\documentclass[12pt]{article}
\usepackage{graphicx}
\usepackage[latin1]{inputenc}
\usepackage{sbc-template}
\usepackage{url}
\usepackage{psfrag}
\usepackage{amsmath}
\usepackage{amsfonts}
\usepackage{amssymb}

\newcommand{\pr}[1]{\protect{#1}}
\newcommand{\ket}[1]{|#1\rangle}
\newcommand{\pket}[1]{\parallel\hskip-1mm #1\rangle}

\newcommand{\pbra}[1]{\langle#1\hskip-1mm\parallel}

\newcommand{\scalar}[2]{\langle#1|#2\rangle}

\newcommand{\op}[1]{|#1\rangle\langle#1|}

\newcommand{\opp}[2]{|#1\rangle\langle#2|}

\newcommand{\intk}{\int_{-\pi}^\pi \frac{dk}{2\pi}}

\title{Long-time entanglement in the Quantum Walk}

\author{Gonzalo Abal\inst{1}, Raul Donangelo\inst{2}, Hugo Fort\inst{3}}

\address{Instituto de Física,  Facultad de Ingeniería, \\
Universidad de la República, C.C. 30, C.P. 11000, Montevideo, Uruguay
\nextinstitute Instituto de Física, Universidade Federal do Rio de Janeiro (UFRJ), \\
C.P. 68528, Rio de Janeiro 21941-972, Brazil
\nextinstitute Instituto de Física, Facultad de Ciencias. \\
Universidad de la República, C.P. 11400, Montevideo, Uruguay
\email{abal@fing.edu.uy, donangel@if.ufrj.br, hugo@fisica.edu.uy}}

\begin{document}
\maketitle

\begin{abstract}
The coin-position entanglement generated by the evolution operator
of a discrete--time quantum walk is quantified, using the von
Neumann entropy of the reduced density operator (entropy of
entanglement). In the case of a single walker, the entropy of entanglement converges, in the
long time limit, to a well defined value which depends on the initial
state. Exact expressions are obtained for local and non-local
initial conditions. We also discuss the asymptotic bi-partite entanglement generated by non-separable
coin operations for two coherent quantum walkers. In this case, the entropy of entanglement is observed to increase logarithmically with time.
\end{abstract}


\section{Introduction}
\label{sec:intro}

The discrete-time Quantum Walk (QW), first introduced in by
Aharonov, Davidovich and Zagury \cite{Aharonov93}, is a quantum
analog for the classical random walk where the classical coin
flipping is replaced by a unitary operation in a one-qubit Hilbert
space. Quantum walks in several topologies \cite{Kempe03} have been
used as the basis for optimal quantum search algorithms
\cite{Shenvi,Childs} which take advantage of quantum entanglement
and parallelism. Entanglement is an essential resource in many
quantum information processing protocols, such as teleportation,
secure key distribution or even Shor's factorization algorithm
\cite{Kendon04}. For pure bi-partite states, it can be quantified using the von
Neumann entropy of the reduced density operator (henceforth, the
entropy of entanglement $S_E$).

The evolution operator of a QW generates quantum correlations between the ``coin" and position degrees of freedom.
This entanglement can be quantified in the long time limit \cite{Abal06} where exact expressions for $S_E$ can be obtained, for given initial conditions.
When two quantum walkers are considered, different kinds of bi-partite
entanglement may appear. We shall discuss this situation and include
some preliminary results for QWs with non-separable coin operators.

This work is organized as follows. In Section~\ref{sec:1pQW}, we
briefly review the discrete-time quantum walk on the line and define
the entropy of entanglement. The dependence of the asymptotic
entanglement for (i) localized initial conditions and (ii) non-local
initial conditions spanned by the position eigenstates $\ket{\pm 1}$
are discussed. Section~\ref{sec:2pQW} is devoted to the case of
two quantum walkers. Finally, in Section~\ref{sec:conc} we summarize
our conclusions and discuss future developments.

\section{Quantum walk on the line}
\label{sec:1pQW}

A step of the QW is a conditional translation between
discrete sites on a line. The Hilbert space is the tensor product of
two subspaces, ${\cal H}={\cal H}_P\otimes{\cal H}_C$. The first
part, ${\cal H}_P $, is a quantum register spanned by the eigenstates
of the position operator,
$X\hskip-1mm\pket{x}=x\hskip-1mm\pket{x}$, with $x$ an integer label. The second part, ${\cal
H}_C$, is a single-qubit ``coin" subspace spanned by two orthonormal
states denoted $\{\ket{0}, \ket{1}\}$. To avoid confusion, we use
the symbol $\pket{\cdot}$ to indicate states in $\cal{H}_P$. A
generic state for the walker is
\begin{equation}
\ket{\Psi}=\sum_{x=-\infty}^\infty \pket{x}\otimes\left[ a_x\ket{0}
+ b_x\ket{1}\right] \label{sp-wv}
\end{equation}
in terms of complex coefficients satisfying $\sum_x
|a_x^2|+|b_x|^2=1$ (in what follows, the summation limits are left
implicit).

The evolution of the walk is described by the unitary operator
\begin{equation}
U=S\cdot\left(I_P\otimes U_C\right) \label{evol1}
\end{equation}
where $U_C$ is a unitary operation in ${\cal H}_C$ and $I_P$ is the
identity in ${\cal H}_P$. A convenient choice for $U_C$ is a
Hadamard operation, defined by
$\protect{H\ket{0}=\left(\ket{0}+\ket{1} \right)/\sqrt{2}}$ and
$\protect{H\ket{1}=\left(\ket{0}-\ket{1} \right)/\sqrt{2}}$. When
$U_C=H$ one refers to the process as a Hadamard walk. The shift
operator
\begin{equation}
S=\sum_x \left\{\pket{x+1}\pbra{x}\otimes\op{0} +
\pket{x-1}\pbra{x}\otimes\op{1}\right\} \label{shift-x}
\end{equation}
conditionally changes the position one step to the right for
$\ket{0}$ or to the left for $\ket{1}$. This conditional shift entangles the coin and
position of the quantum walker.

In terms of the density operator, $\rho=\op{\Psi}$, the evolution is
given by  $\protect{\rho(t)=U^t\rho(0)\,U^{\dagger\,t}}$ where the
non-negative integer $t$ counts the discrete time steps that have
been taken. The probability distribution for finding the walker at
site $x$ at time $t$ is
$\protect{P(x,t)=tr(\rho P_x)=|a_x|^2+|b_x|^2}$, where $P_x=\pket{x}\pbra{x}\otimes I_c$. The
variance of this distribution increases quadratically with time as
opposed to the classical random walk, in which the increase is only
linear. This property depends on quantum coherence and is lost in
the presence of noise \cite{qw-markov, deco}.

For pure, bi-partite states, entanglement can be quantified by the
Entropy of Entanglement \cite{Bennet96,Myhr}, defined as the von
Neumann entropy of the reduced density operator. If the partial
trace is conveniently taken over the position subspace,
$\rho_c=tr_x(\rho)$, and
\begin{equation}
S_E\equiv -tr(\rho_c\log_2\rho_c).\label{SC}
\end{equation}
This quantity is zero for a product state, unity for a maximally
entangled state and it is invariant under LOCC (local operations with
classical communication) \cite{Vedral,SM95}.

In an early paper, Nayak and Vishwanath \cite{Nayak} have shown that
Fourier analysis may be used to obtain long-time asymptotic
expressions for the amplitudes $a_x(t)$ and $b_x(t)$, for given
initial conditions. We used this approach to characterize the
asymptotic entanglement of the QW in \cite{Abal06}. In the rest of
this Section, we summarize and extend those results.

We start by defining the dual space, $\tilde{\cal H}_k$, spanned by the Fourier
transformed kets $\ket{k}=\sum_x e^{ikx}\ket{x}$, with real
wavenumbers $k\in [-\pi,\pi]$. In this representation
the state vector is
\begin{equation}
\ket{\Psi}=\intk\ket{k}\otimes\left[\tilde a_k\ket{0} + \tilde b_k\ket{1}\right].\label{wv-k}
\end{equation}
of the QW, ,

These amplitudes are related to the position amplitudes in eq.~(\ref{sp-wv}) by
$\protect{\tilde a_k = \sum_x e^{-ikx}a_x}$ and $\protect{\tilde b_k
= \sum_x e^{-ikx}b_x}$, respectively. Since the step size is
constant, the shift operator defined in eq.~(\ref{shift-x})
\begin{equation}
U_k=\frac{1}{\sqrt{2}}\left(
\begin{array}{cc}
e^{-ik} & e^{-ik} \\
e^{ik} & -e^{ik}
\end{array} \right).
\label{k-evol}
\end{equation}
is diagonal in $k$. Then, if $\ket{\Phi_k}$ is the spinor $(\tilde a_k, \tilde b_k)^T$, the time evolution of an initial state can
be expressed as
\begin{equation}\label{k-evol1}
\ket{\Phi_k(t)}= U_k^t\ket{\Phi_k(0)}= e^{-i\omega_k
t}\scalar{\varphi_k^{(1)}}{\Phi_k(0)}\;\ket{\varphi_k^{(1)}}+ (-1)^t
e^{i\omega_k
t}\scalar{\varphi_k^{(2)}}{\Phi_k(0)}\;\ket{\varphi_k^{(2)}}.
\end{equation}
where $\ket{\varphi_k^{(1,2)}}$ are the eigenvectors and  $\pm
e^{\mp i\omega_k}$ the eigenvalues of $U_k$ and the angle
$\protect{\omega_k\in[-\pi/2,\pi/2]}$ is defined by
$\protect{\sin\omega_k\equiv\sin k/\sqrt{2}}$. In principle,
eq.~(\ref{k-evol1}) may be transformed back to position space and
the probability distribution $P(x,t)$ obtained implicitly in terms
of complicated integrals which, for arbitrary times, can only be
done numerically. However, in the long time limit, approximate
stationary phase methods can be used to evaluate these integrals for
given initial conditions \cite{Nayak}. We may avoid these technical
difficulties, because the asymptotic entanglement $S_E$ may be
obtained directly from eq.~(\ref{k-evol1}) and there is no need to
transform back to position space.

The entropy of entanglement of a quantum walker can be obtained
after diagonalisation of the reduced density operator
\begin{equation}
\rho_c=tr_x(\rho)=\left(
\begin{array}{cc}
 A& B \\
 B^*&C
\end{array}\right), \label{rho_C}
\end{equation}
where
\begin{eqnarray}
A&\equiv&\sum_x|a_x|^2=\intk |\tilde a_k|^2\nonumber\\
B&\equiv&\sum_x a_xb_x^*=\intk \tilde a_k\tilde b_k^*\label{coefs}\\
C&\equiv&\sum_x|b_x|^2=\intk |\tilde b_k|^2.\nonumber
\end{eqnarray}
Normalization requires that tr$(\rho)=A+C=1$. In terms of
\begin{equation}
\Delta\equiv AC-|B|^2\label{disc}
\end{equation}
the real, positive eigenvalues of $\rho_c$ are $r=\frac12\left[1+\sqrt{1-4\Delta} \right] $ and $1-r$.
The entropy of entanglement is obtained from $\protect{S_E=-r\log_2 r -(1-r)\log_2 (1-r)}$.
The units for bi-partite entanglement are e-bits, or entanglement bits, where one e-bit is the amount of entanglement contained in a Bell pair.
The determinant $\Delta\in[0,1/4]$ contains all the information required to quantify the coin-position entanglement in the QW. The greater the value of $\Delta$, the greater the entanglement.

\subsection{Local initial coins}

\begin{figure}[t]
\centering
\includegraphics[scale=0.4,angle=-90]{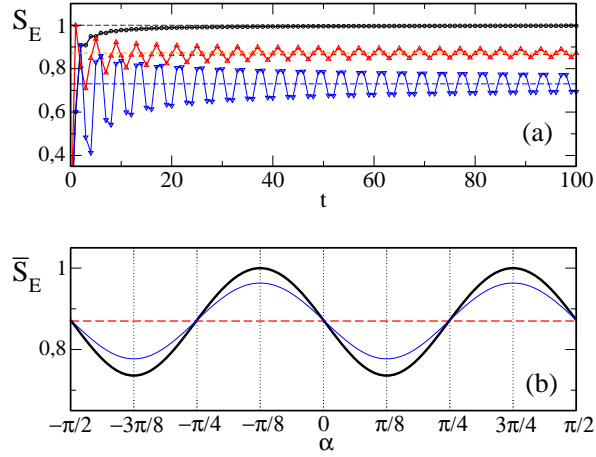}
\caption{(color online) Entanglement vs discrete time for localized initial states
with coin states $\ket{\chi}$ parametrized in eq.~(\ref{ic1}); (a)  time evolution
of the entropy of entanglement $S_E$ for $\alpha=-\pi/8$ and
$\beta=\pi$ (black, full asymptotic entanglement), $\beta=\pi/2$
(red, intermediate asymptotic entanglement) and $\beta=0$, (blue,
minimimum asymptotic entanglement); (b) asymptotic entropy of
entanglement $\bar S_E$ vs $\alpha$ from eq.~(\ref{disc2}) for
$\beta=0$ (thick black line) $\beta=\pi/4$, (thin blue line) and
$\beta=\pm\pi/2$ (dashed red line). } \label{fig:SE_local}
\end{figure}

Consider a localized initial state ($x=0$, without loss of generality) with an arbitrary coin parametrized in
terms of two real angles $\alpha\in[-\pi/2,\pi/2]$ and  $\beta\in [-\pi,\pi]$,
\begin{equation}
\ket{\chi}\equiv\frac{1}{\sqrt{2}}\left[ \cos\alpha\ket{0}+e^{i\beta}\sin\alpha\ket{1}\right] .
\label{ic1}
\end{equation}
In this case, the Fourier-transformed coefficients appearing in eq.~(\ref{wv-k}) are $\pr{\tilde a_0(0)=\cos\alpha}$, $\pr{\tilde b_0(0)=e^{i\beta}\sin\alpha}$ and  $\pr{\tilde a_k(0)=\tilde b_k(0)=0}$ for $k\ne 0$.
The integrands in eqs.~(\ref{coefs}) can be calculated from eq.~(\ref{k-evol1}) in terms of $\alpha$ and $\beta$.
For long times, $t\gg 1$, the $k$-averages in eqs.~(\ref{coefs}) are time-independent and the dependence of $\Delta$ on the initial coin is
\begin{equation}
\bar \Delta\equiv\lim_{t\gg 1}\Delta=\Delta_0-2b_1^2\cos\beta\sin(4\alpha),\label{disc2}
\end{equation}
with $\Delta_0=(\sqrt{2}-1)/2$ and $b_1=(2-\sqrt{2})/4$. The details
of the calculation leading to this expression can be found in Appendix A of Ref.~\cite{Abal06} and will not be reproduced here.
In panel (a) of Fig.~\ref{fig:SE_local} the time evolution of the entropy of entaglement is shown for fixed $\alpha=-\pi/8$ and $\beta=\pi,\pi/2,0$ for which the asymptotic entanglement is maximum, intermediate or minimum respectively. In particular, the intermediate asymptotic entanglement level,
$\bar S_E\approx 0.872$, has been erroneously reported as generic (i.e. resulting from \textit{all} initial coins) in Ref.~\cite{Carneiro}. In fact, as follows from expression (9), it is obtained for arbitrary $\alpha$ only if $\beta=\pi/2$. Also apparent from Fig.~\ref{fig:SE_local} is the fact that lower entanglement levels result in larger oscillations and have a slower convergence to their asymptotic values. Panel (b) of this figure shows the variation of the asymptotic entanglement with $\alpha$ for three values of the relative phase $\beta$. For localized initial coins the minimum asymptotic entanglement is $\bar S_E\approx 0.736$. However, as we discuss next, if non-local initial conditions are considered, arbitrary low values for asymptotic entanglement may be obtained.

\subsection{Non-local initial coins}

\begin{figure}
\psfrag{theta}{$\quad\theta$}\psfrag{phi}{~\small$\varphi$}
\psfrag{mpi/2}{$-\pi/2$}\psfrag{mpi/4}{$-\pi/4$}
\psfrag{pi/2}{$\pi/2$}\psfrag{pi/4}{$\pi/4$}
\psfrag{pi}{$~\pi$}\psfrag{mpi}{\hskip-2mm $-\pi$}
\includegraphics[scale=0.65]{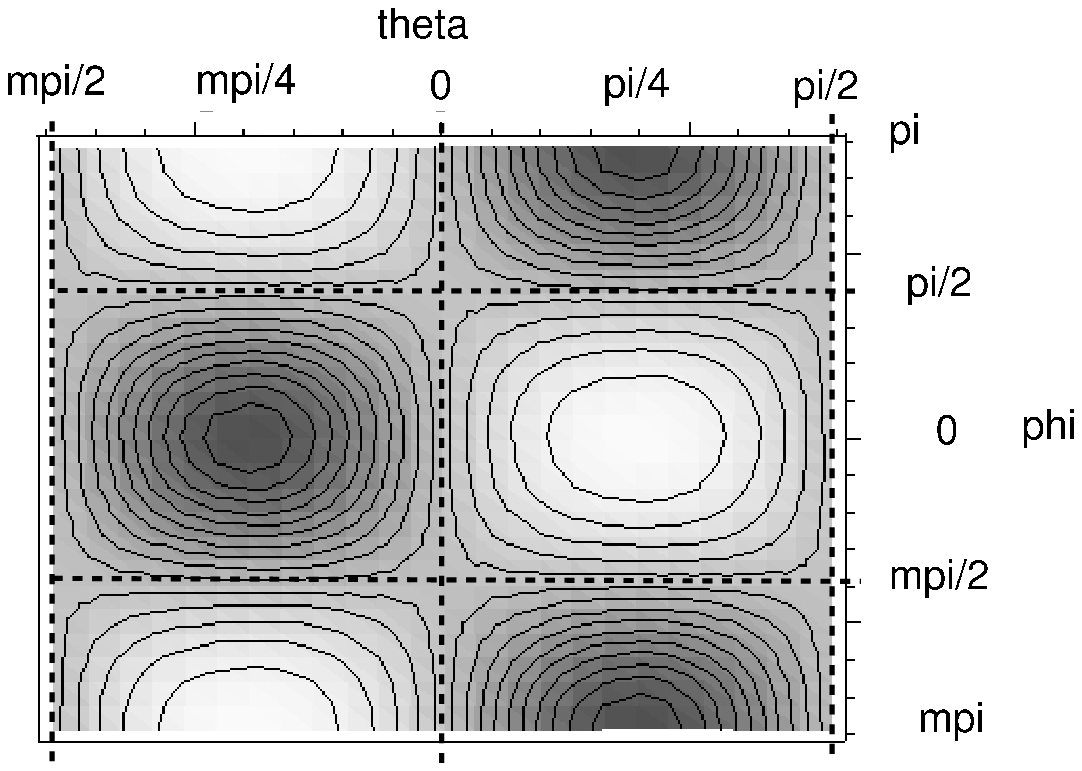}\vskip-6cm\hskip8cm
\includegraphics[scale=0.4,angle=-90]{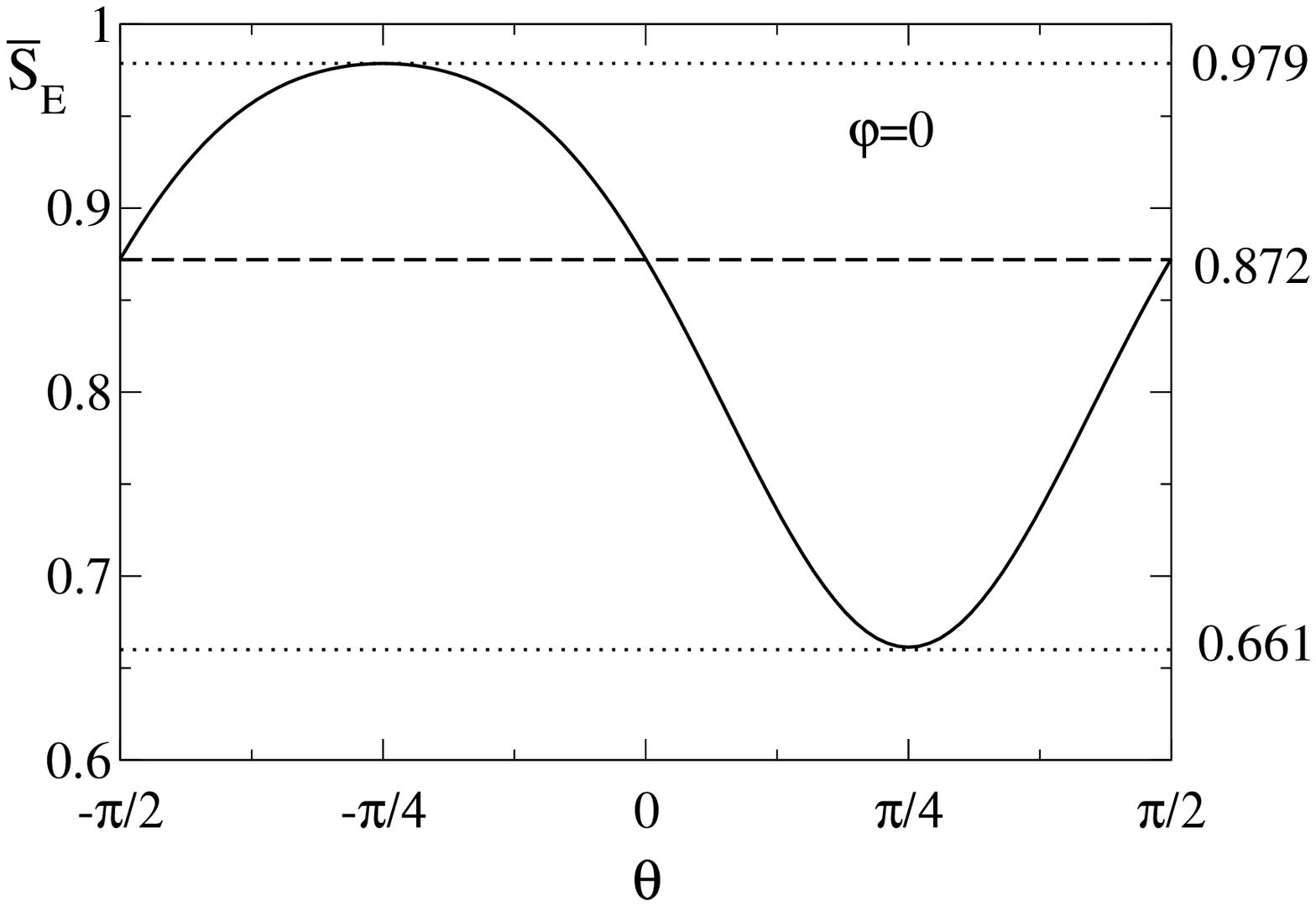}
\caption{(Left) contour plot of the asymptotic entanglement for non-local initial conditions obtained from eq.~(\ref{ex-ev}).
Clear areas indicate maxima and dark areas, the minimum values.
(Right) variation of asymptotic entanglement with non-locality $\theta$ for the particular relative phase $\varphi=0$.} \label{fig:sc4-contour}
\end{figure}

In order to explore the effects of non-locality in the initial
conditions, let us consider a generic ket in the position subspace
spanned by the eigenkets $\pket{\pm 1}$,
\begin{equation}
\ket{\Psi(\theta,\varphi)}=\left[\cos\theta\pket{-1}+e^{-i\varphi}\sin\theta\pket{1}\right]
\otimes\ket{\chi_0}.\label{Psi_gen}
\end{equation}
where the parameters $\protect{\theta\in[-\pi/2,\pi/2]}$ and
$\protect{\varphi\in[-\pi,\pi]}$ are real angles. The initial coin
is fixed at
\begin{equation}\label{icoin}
\ket{\chi_0}=\frac{\ket{0}+i\ket{1}}{\sqrt{2}}.
\end{equation}
Notice that in this case, localized states ($\theta=0,\pm\pi/2$) result in the intermediate entaglement $\bar S_E\approx 0.872$.  With these initial conditions, exact asymptotic expressions for the coefficients defined in eqs.~(\ref{coefs}) can be obtained (again, for the details see
Ref.~\cite{Abal06}) leading to the eigenvalues
\begin{equation}
\bar r_{1,2}=\frac12 \pm \left[\left(B_0-B'\sin 2\theta\cos\varphi \right)^2 +
B_+^2\sin^2 2\theta\sin^2\varphi\right]^{1/2}\label{ex-ev}
\end{equation}
where $B_0=(\sqrt{2}-1)/2$, $B'=(3\sqrt{2}-4)/2$ and $B_+=(\sqrt{2}-1)^2/2$. From this expression, the asymptotic
entropy of entanglement $\bar S_E(\theta,\varphi)$ can be evaluated exactly. The left panel of Fig.~\ref{fig:sc4-contour}
shows a contour plot of this surface. For the initial conditions $\left(\pket{-1}\pm\pket{1}\right)/\sqrt{2}$,
the asymptotic entanglement is maximum $\protect{\bar S_E\approx
0.979}$ or minimum $\protect{\bar S_E\approx 0.661}$, respectively.
The vertical dashed lines indicate initially localized position
eigenstates and the horizontal dashed lines indicate initial
position eigenstates with relative phase $\varphi=\pm\pi/2$. In both cases,
the intermediate asymptotic entanglement $\protect{\bar S_E\approx
0.872}$ results. The right panel of Fig.~\ref{fig:sc4-contour} shows the variation of $\bar S_E$ with non-locality $\theta$ in the particular case when the relative phase is fixed at $\varphi=0$.

The minimum entanglement that can be obtained from non-local initial
conditions in this subspace ($\protect{\bar S_E\approx 0.661}$) is
lower than the one attainable from local initial conditions ($\bar
S_E\approx 0.736$, see Fig.~\ref{fig:SE_local}). This raises a question about if
further non-locality will result in even lower asymptotic entanglement levels.
To illustrate the point, consider an initial Gaussian wave packet
with a characteristic spread $\sigma\gg 1$  in position space, with
the same coin state $\ket{\chi_0}$ as before. In this case, the
Fourier transformed coefficients, $\protect{\tilde a_k(0)\propto
\sqrt{\sigma}\,e^{-k^2\sigma^2/2}}$, describe a localized
state in $k$-space. In fact, $\protect{\lim_{\sigma\rightarrow\infty} |\tilde
a_k(0)|^2=2\pi\delta(k)}$, where $\delta(k)$ is Dirac's delta
function. In this limit, the eigenvalues of $\rho_c$ reduce to $0$ and $1$ and the corresponding
asymptotic entropy of entanglement becomes vanishingly small. Thus, for a particular uniform initial distribution in position space,
a product state results in the long time limit if the
appropriate relative phases are chosen.

\section{Two entangled walkers}
\label{sec:2pQW}

Entanglement  in two-particle quantum walks \cite{Carneiro, Mackay, Omar, Tregenna} has not been fully
characterized yet.  The Hilbert space for two walkers is just the tensor product of two
one-particle spaces, ${\cal H}_{AB}={\cal H}_A\otimes{\cal H}_B$,
where both ${\cal H}_A$ and  ${\cal H}_B$ are isomorphic to the
one-particle space ${\cal H}={\cal H}_C\otimes{\cal H}_P$ described
in Section~\ref{sec:1pQW}. We label the positions of the walkers
with pairs of integers $(x,y)$, so that a generic two-particle pure
state $\ket{\Psi}$ is
\begin{equation}\label{2part-wv}
\ket{\Psi}=\sum_{x,y}\left\{\alpha_{x,y}\ket{00} +
\beta_{x,y}\ket{01}+ \gamma_{x,y}\ket{10}+
\delta_{x,y}\ket{11}\right\}\otimes\pket{x,y}
\end{equation}
with complex coefficients satisfying the normalization requirement
\begin{equation}\label{2norm}
\sum_{x,y}|\alpha_{x,y}|^2+|\beta_{x,y}|^2+|\gamma_{x,y}|^2+|\delta_{x,y}|^2=1.
\end{equation}

The two-particle evolution operator is composed of a
unitary operation $U_C$ in the two-qubit coin subspace ${\cal
H}_C^{\otimes 2}$, followed by a conditional shift $S$ in
position space,
\begin{equation}
U_{AB}=S\cdot\left(I_P\otimes U_C\right)\label{ev-op12}
\end{equation}
where $I_P$ is the identity in the two-particle position subspace
${\cal H}_P^{\otimes 2}$. The shift operator
\begin{eqnarray}
S&=&\sum_{x,y}\left\{\opp{x+1,y+1}{x,y}\otimes\op{00} +
\opp{x+1,y-1}{x,y}\otimes\op{01}\right.\nonumber\\
&&\quad\left. +\opp{x-1,y+1}{x,y}\otimes\op{10}
+\opp{x-1,y-1}{x,y}\otimes\op{11}\right\}\label{shift-2p}
\end{eqnarray}
performs single-step conditional displacements.  An initial two-particle pure state, characterized by a density
operator $\rho(0)=\op{\Psi(0)}$, evolves to
\begin{equation}
\rho (t)=U_{AB}^t\,\rho(0)\left(U_{AB}^\dagger\right)^t\label{evol-12}
\end{equation}
after $t$ time steps. The resulting joint probability distribution
for finding walker $A$ at site $x$ and walker $B$ at site $y$, is
\begin{equation}
P_{AB}(x,y; t)=tr_{C}(\rho)=
|\alpha_{x,y}|^2+|\beta_{x,y}|^2+|\gamma_{x,y}|^2+|\delta_{x,y}|^2.
\label{dist12}
\end{equation}
Note that, if $\rho$ describes a pure, separable state (i.e. $\rho=\rho_A\otimes\rho_B$), the joint
distribution is a product of the two single particle distributions,
$\protect{P_{AB}(x,y; t)=P_A(x,t)P_B(y,t)}$. In the generic case,
$\rho$ describes an entangled state and a measurement of the
position of one walker will affect the position of the other. In this initial work, we restrict consideration to pure states, since the quantification of entanglement in mixed states is more involved.

\begin{figure}
\centering
\includegraphics[width=4.5cm,angle=-90]{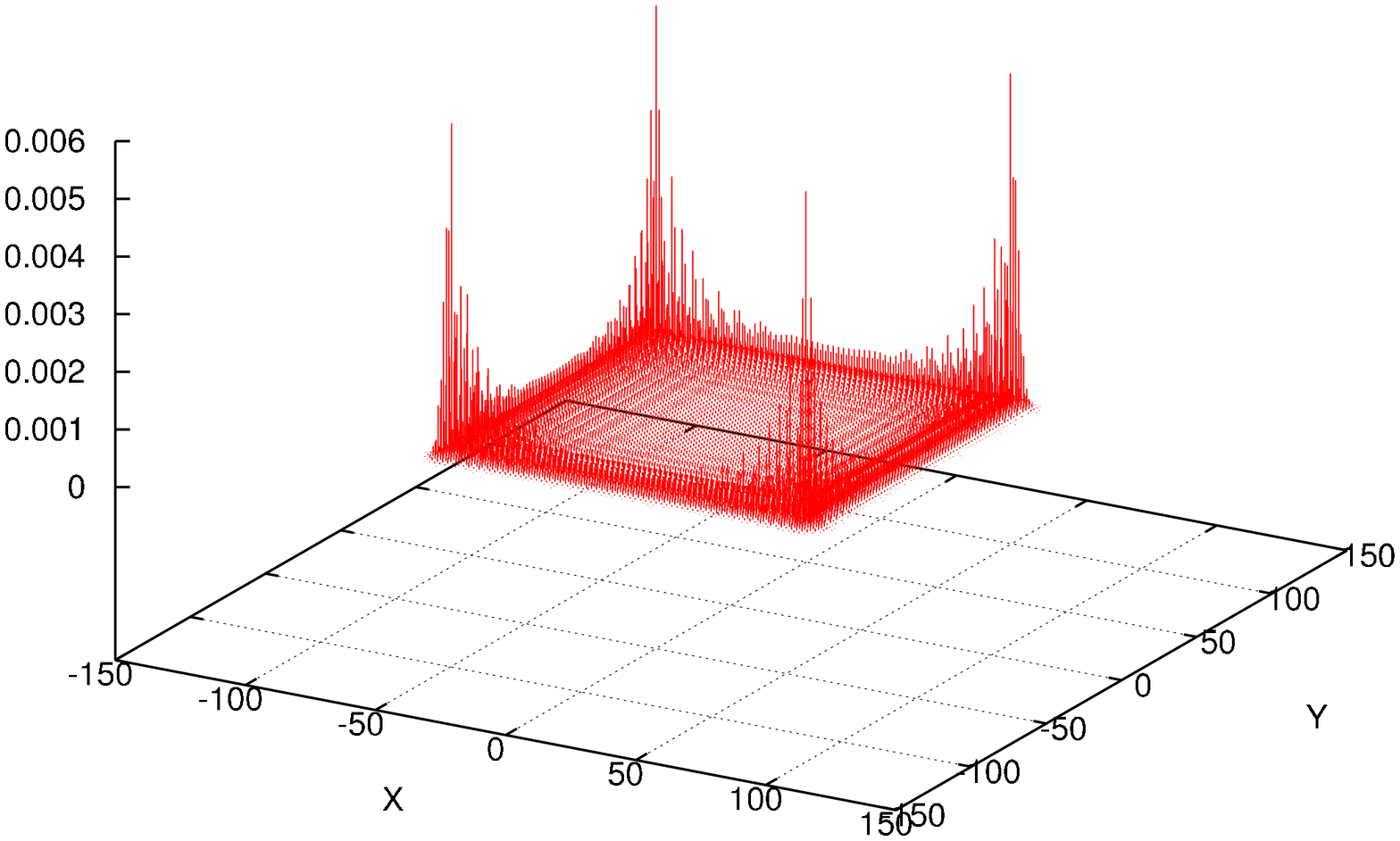}
\includegraphics[width=4.5cm,angle=-90]{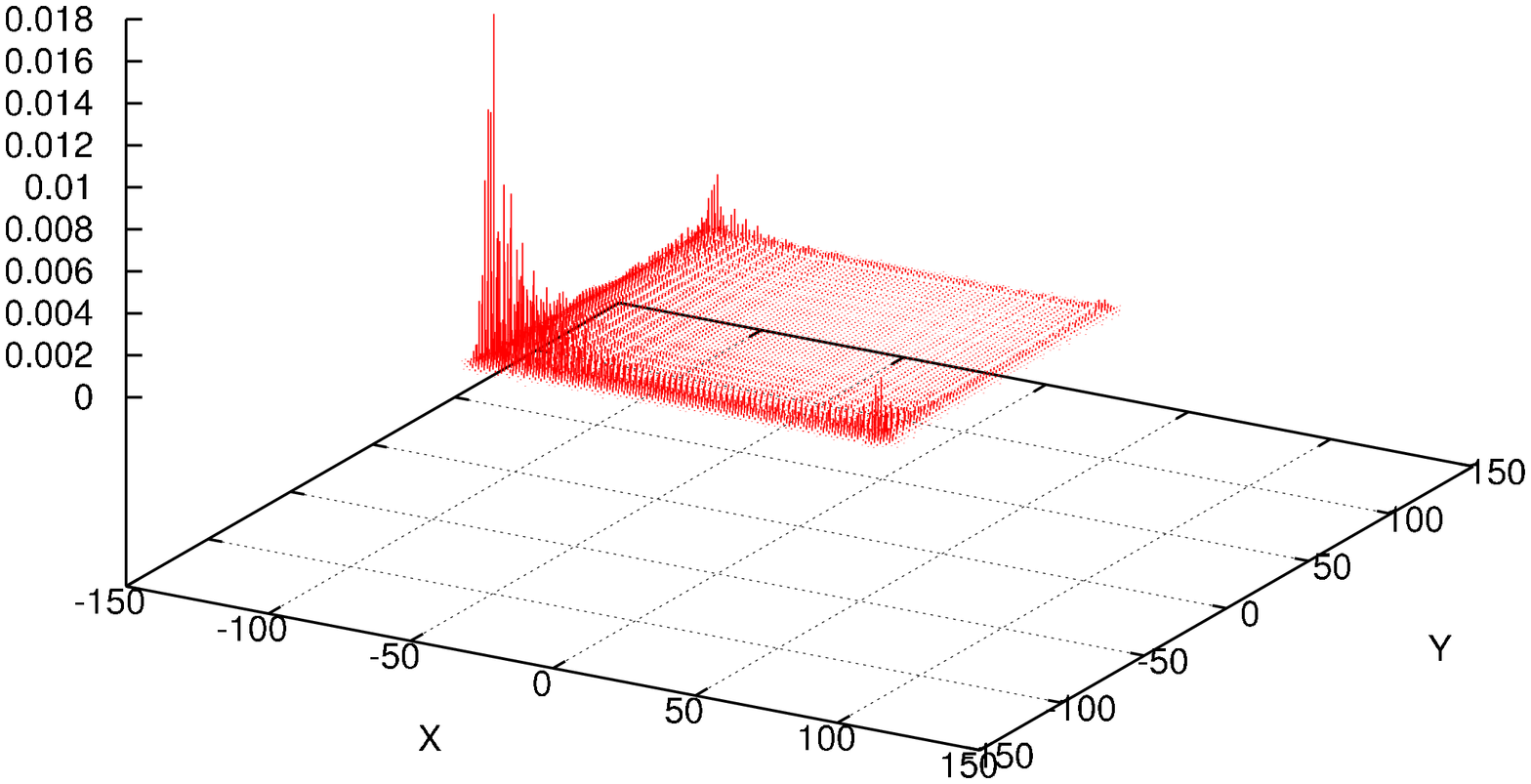}
\caption{DRAFT Probability distribution $P(x,y)$ after $t=100$ steps with the separable Hadamard coin, eq.~(\ref{H2-coin}).  (Left) Initial coins  $\ket{\chi_1}$, eq. ~(\ref{s1}), which result in a symmetrical evolution, i.e. $P_A(x,t)=P_A(-x,t)$ and $P_B(y,t)=P_B(-y,t)$;  (Right) Initial coins   $\ket{\chi_2}$, eq. ~(\ref{s2}). In both cases, there is no entanglement between A and B.}
\label{fig:hadamard-dist}
\end{figure}

\subsection{Separable coin operations}

In the simplest case, the coin operation $U_C$ in eq.~(\ref{ev-op12}) may be separable,
$$
U_C=U_A\otimes U_B,
$$
where $U_A$ and $U_B$ are local unitary operators in ${\cal H}_C$. For these coin operations, entanglement between
subspaces ${\cal H}_A$ and ${\cal H}_B$ can arise only from the choice of initial conditions and it is left unchanged by the
evolution. As a simple example, let us consider the two--particle Hadamard
walk with
\begin{equation}
U_C=H\otimes H=\frac12 \left(
\begin{array}{cccc}
 ~1 &~1  &~1&~1  \\
 ~1& -1 &~1 &-1 \\
 ~1& ~1 &-1 &-1\\
 ~1& -1 &-1 &~1
\end{array}\right).\label{H2-coin}
\end{equation}
Figure~\ref{fig:hadamard-dist} shows the probability distribution obtained after $t=100$ steps, from two initial initial conditions localized at the origin with the balanced coins
\begin{eqnarray}
\ket{\chi_1}&=&\frac12\left[\ket{00}+i\ket{01}+i\ket{10}-\ket{11}\right]\label{s1}\\
\ket{\chi_2}&=&\frac12\left[\ket{00}-\ket{01}-\ket{10}+\ket{11}\right].\label{s2}
\end{eqnarray}
Note that both coins are separable so that in this case there is no entanglement between the two particles.

\begin{figure}
\centering
\includegraphics[width=4.5cm,angle=-90]{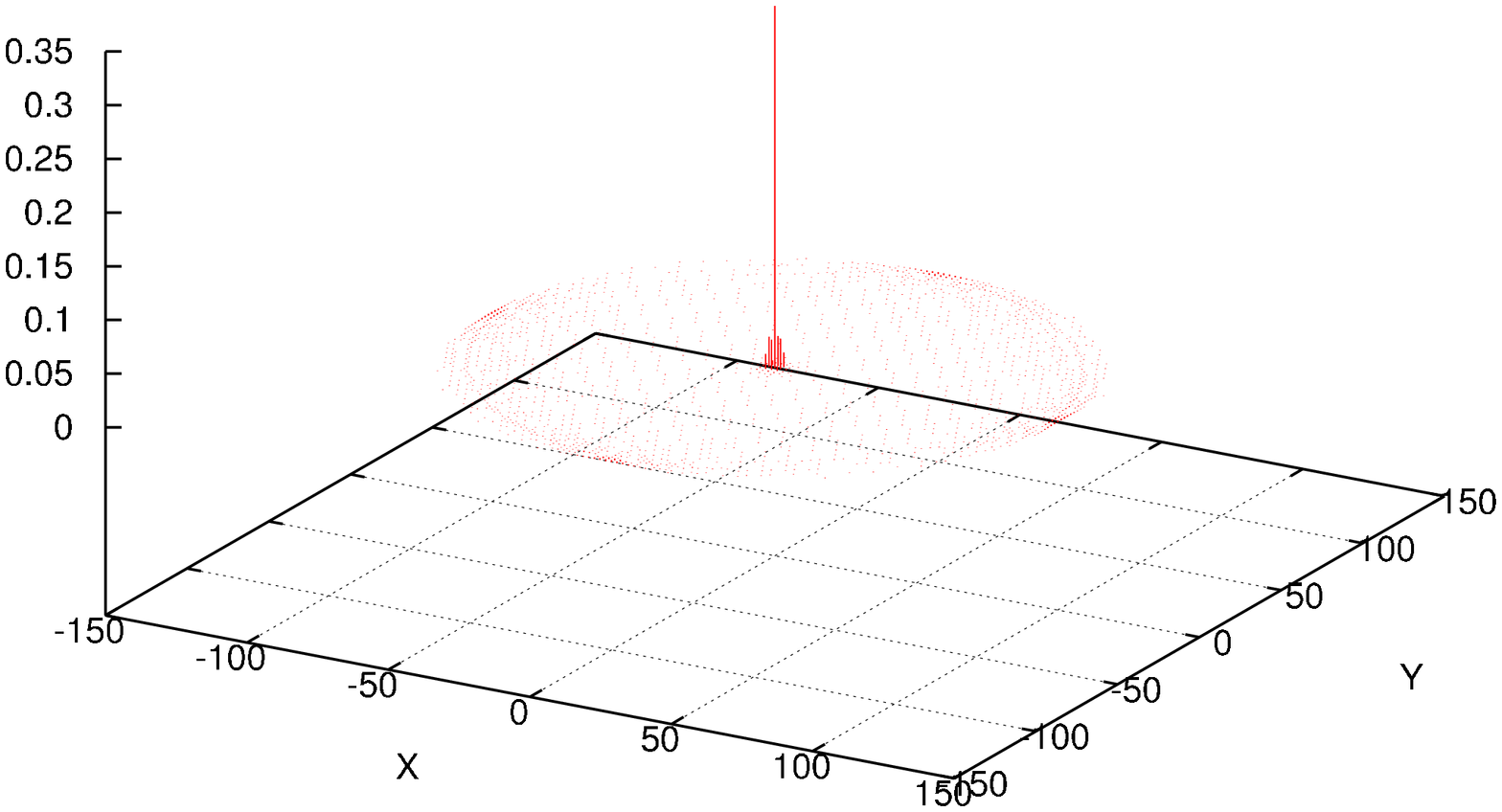}~\includegraphics[width=4.5cm,angle=-90]{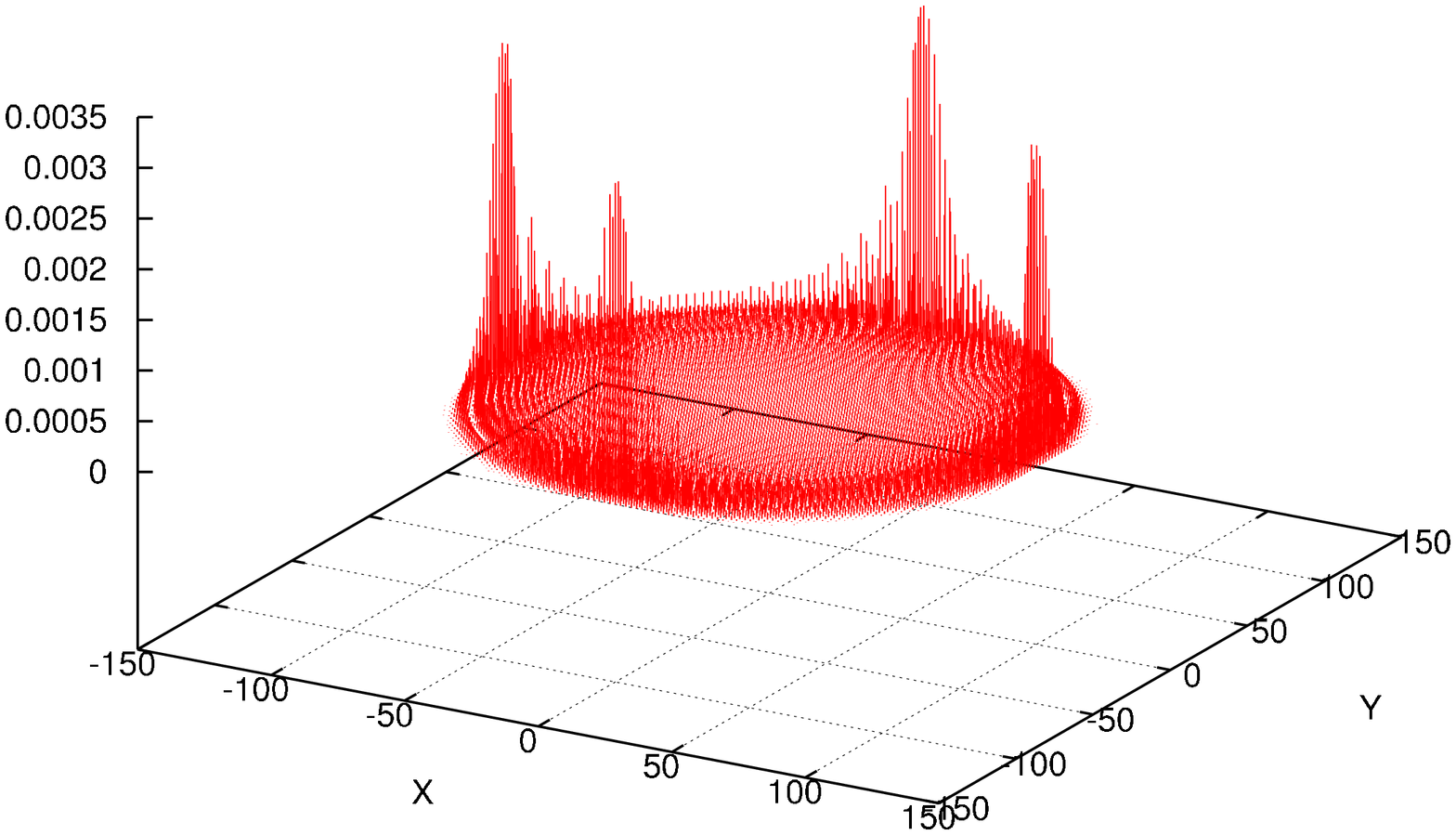}
\caption{DRAFT Probability distribution $P(x,y)$ after $t=100$ steps with Grover' s coin, eq.~(\ref{Grover4-coin}).  (Left) Initial coins  $\ket{\chi_1}$, eq. ~(\ref{s1}), which results in a distribution which remains highly peaked at the origin;  (Right) Initial coins  $\ket{\chi_2}$, eq. ~(\ref{s2}), which results in maximum spread.}
\label{fig:grover-dist}
\end{figure}

\subsection{Non-separable coin operations}

A more interesting situation is the case of non-separable coins which
may change the entanglement between the A and B subspaces. Now entanglement may be introduced by
the initial condition or by the coin operation. There are now several kinds of entanglement, since the shift operation still entangles the coin and position degrees of freedom as described in the previous section. We shall consider the entanglement between both particles, generated by the coin operation $U_C$, thus we use unentangled initial coins from eqs. (\ref{s1}) and (\ref{s2}).

\begin{figure}[ht]
\hskip2cm\includegraphics[width=3cm]{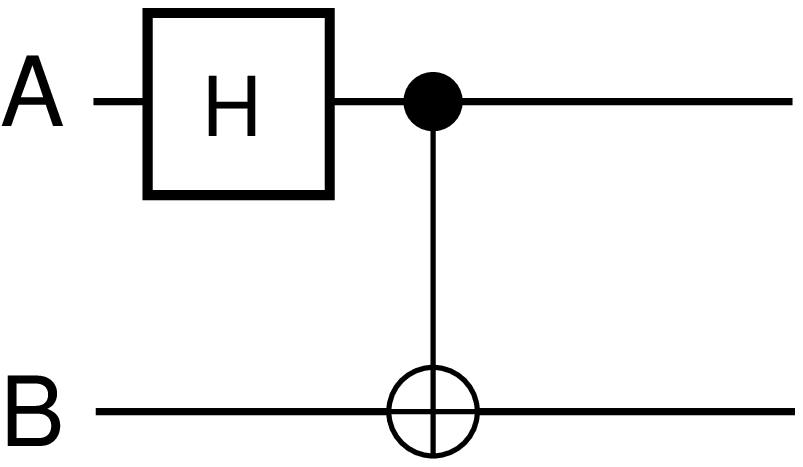}\vskip-4cm\hskip6cm \includegraphics[width=5cm,angle=-90]{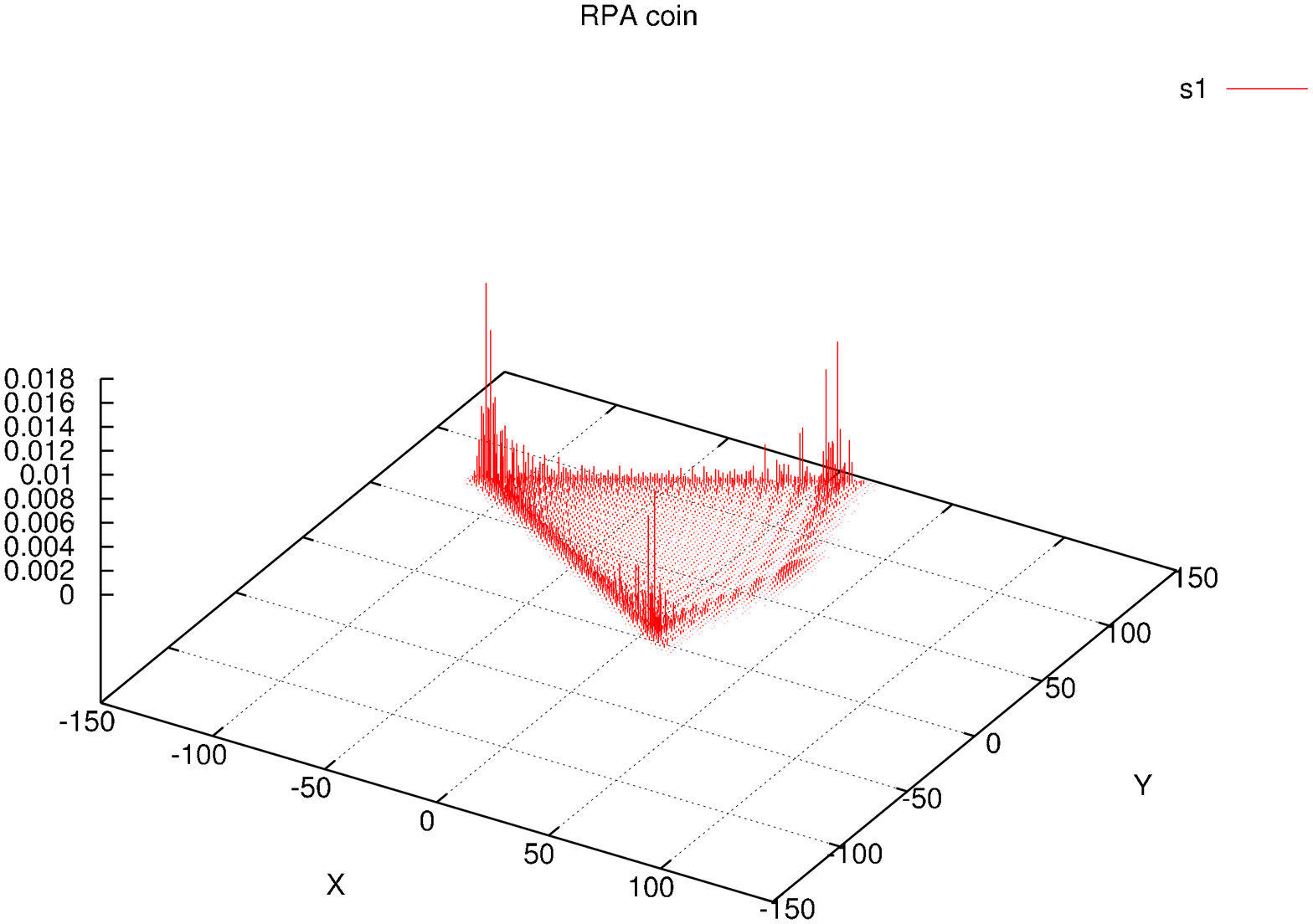}
\caption{(Left)  Schematic diagram showing the action of the coin operation (\ref{RP-coin}) in two-qubit space;  (Right) Probability distribution $P(x,y)$ after $t=100$ steps  with the coin operaton defined in eq.~(\ref{RP-coin}) corresponding to a Random-Pavlov quantum game. The initial state is localized with coins $\ket{\chi_1}$, eq. ~(\ref{s1}).}
\label{fig:rpa-dist}
\end{figure}

As a first example of non-separable coin operator, consider Grover's coin
\begin{equation}
G=\frac12 \left(
\begin{array}{cccc}
-1 &1  &~1&~1  \\
 ~1& -1 &~1 &~1 \\
 ~1& ~1 &-1 &~1\\
 ~1& ~1 &~1 &-1
\end{array}\right)\label{Grover4-coin}
\end{equation}
which plays a central role in Grover's search algorithm \cite{Grover1}. The probability distribution corresponding to this coin operation remains strongly peaked at the origin for most initial coins, see Fig.~\ref{fig:grover-dist}, left panel. However, for the specific initial coin $\ket{\chi_2}$, eq.~(\ref{s2}),  the distribution has a maximum spread \cite{Tregenna}, see right panel.

\begin{figure}[ht]
\centering
\includegraphics[scale=0.6,angle=-90]{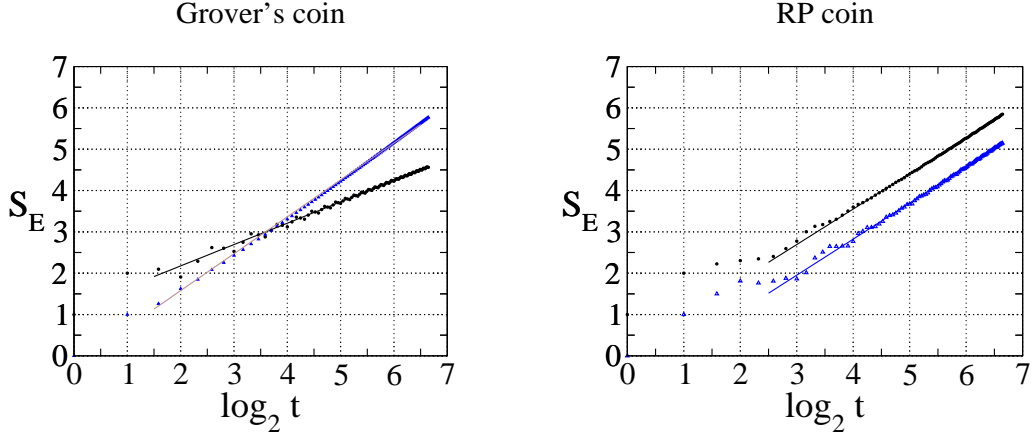}
\caption{(color online) Entropy of entaglement $S_E$ vs $log_2(t)$ for Grover's coin operation (left panel) and RP coin operation (right panel). Two initial states corresponding to localized states with coins $\ket{\chi_1}$ (black circles) and $\ket{\chi_2}$ (blue triangles) are shown. The straight lines are the linear regression fits as discussed in the text. }
\label{fig:ent-all}
\end{figure}

Another example of considerable interest is provided by the coin operations associated to quantum strategies in bi-partite quantum games \cite{Abal-qg}. In particular, the coin operation
\begin{equation}
U_C=CNOT\cdot(H\otimes I)=\frac{1}{\sqrt{2}} \left(
\begin{array}{cccc}
~1 &~0  &~1&~0  \\
 ~0& ~1 &~0 &~1 \\
 ~0& ~1 &~0 &-1\\
 ~1& ~0 &-1 &~0
\end{array}\right)\label{RP-coin}
\end{equation}
describes a quantum game (RP) in which agent A implements a
Random-like strategy, represented by a Hadamard operation $H$ and B
responds with a particular Pavlovian strategy (using a CNOT gate).
Note that this coin operation, described by the circuit in
Fig.~\ref{fig:rpa-dist} (left),  generates the Bell states from the
computational basis states.  The resulting probability distribution
(right panel) has a triangular form in $(x,y)$ space.

The time dependence of the bi-partite entanglement generated by the
Grover and RP coin operations, quantified with the entropy of
entanglement $S_E$,  is shown in Fig.~\ref{fig:ent-all}. It
increases logarithmically with the number of iterations, namely
$S_E\sim log_2(t^c)$. This may be associated to the fact that more
sites on the plane become occupied and more position eigenstates
become entangled as time increases. We estimated the constant $c$
using a linear regression. For Grover's coin operation with initial
coin $\ket{\chi_1}$,  which leads to minimum spread, $c= 0.52$. In
the case of the initial coin $\ket{\chi_2}$,  which leads to maximum
spread, the entanglement increases faster and $c= 0.89$.  The RP
coin, while producing a very different spatial distribution,
generates entanglement (for both initial conditions) with $c= 0.87$,
which is a very similar rate to that of Grover's coin in the
maximally spreading case. These linear fits are the straight lines
shown in Fig.~\ref{fig:ent-all}.

\section{Conclusions}
\label {sec:conc}
The evolution operator of the QW generates different kinds of entanglement. The conditional shift operation entangles the coin and position degrees of freedom of each walker. This kind of entanglement has a well defined asymptotic value that depends on the initial conditions. In the case of a Hadamard walk with spatially localized initial conditions, the asymptotic entanglement varies between almost full entanglement and a minimum entanglement of $\bar S_E\approx 0.736$. However, when non-local initial conditions are considered, any asymptotic entanglement level may be obtained.

When two quantum walkers with a non-separable coin operation are considered, the inter-particle entanglement increases at a logarithmic rate. The time dependence of the entanglement generated by Grover's coin and by the RP coin (associated to a particular strategic choice in bi-partite quantum games) has been considered for two initial coins. In most of the cases, the entanglement increases as $S_E\sim \log_2 t^c$ with $c\approx 0.9$. However, for Grover's coin operation  with the initial coin $\ket{\chi_1}$  (which leads to a localized probability distribution), the entanglement increases more slowly, with $c=0.52$. Entanglement is a basic resource in quantum algorithms, and further work is required in order to fully understand its properties in the QW.  The analytical method outlined in Section 2 for the case of a single walker, may be extended to the case of two walkers and this should provide a more profound understanding of the time dependence we have described in Section 4 of this work.

\noindent Akwnowledgements. {\it Work supported by  PEDECIBA and PDT under project 29/84 (Uruguay).  R.D. acknowledges support from CNPq and FAPERJ (Brazil).}

\bibliographystyle{h-physrev} 

\bibliography{qwnloc}
\end{document}